\documentclass[letterpaper, 10 pt, conference]{ieeeconf}

\IEEEoverridecommandlockouts                    

\usepackage{graphicx,amssymb,xcolor,amsmath,cite,url}

\usepackage{amsthm}
\usepackage{comment}
\usepackage{algorithm}
\usepackage{algorithmic}

\makeatletter
\newenvironment{algsubsteps}{\begin{ALC@g}}{\end{ALC@g}}
\makeatother

\usepackage{tikz}
\usetikzlibrary{shapes,arrows,positioning,calc}
\tikzset{
block/.style = {draw, fill=white, rectangle, minimum height=2.5em, minimum width=3em},
tmp/.style = {coordinate},
sum/.style= {draw, fill=white, circle, node distance=1cm},
input/.style = {coordinate},
output/.style= {coordinate},
pinstyle/.style = {pin edge={to-,thin,black}}}

\theoremstyle{remark}
\newtheorem{remark}{Remark}

\title{\LARGE \bf CPU- and GPU-Based Parallelization of the Robust Reference Governor}

\author{Hamid R. Ossareh$^{1}$, William Shayne$^{2}$, Samuel Chevalier$^{1}$
\thanks{This paper is based upon research supported by the U.S. DoE under award number DE-EE0010407. The views expressed herein do not necessarily represent the views of the U.S. DoE or the United States Government.}
\thanks{$^{1}$ Samuel Chevalier and Hamid R. Ossareh are with the Department of Electrical and Biomedical Engineering, The University of Vermont, Burlington, VT, USA. E-mails:
{\tt\small \{samuel.chevalier, hamid.ossareh\}@uvm.edu}}
\thanks{$^{2}$ William Shayne is with the Department of Computer Science, The University of Vermont, Burlington, VT, USA. E-mail: 
\tt\small william.shayne@uvm.edu}
}%

\begin{document}

\maketitle
\thispagestyle{empty}
\pagestyle{empty}

\begin{abstract}
Constraint management is a central challenge in modern control systems. A solution is the Reference Governor (RG), which is an add-on strategy to pre-stabilized feedback control systems to enforce state and input constraints by shaping the reference command. While robust formulations of RG exist for linear systems, their extension to nonlinear systems is often computationally intractable. This paper develops a scenario-based robust RG formulation for nonlinear systems and investigates its parallel implementation on multi-core CPUs and CUDA-enabled GPUs. We analyze the computational structure of the algorithm, identify parallelization opportunities, and implement the resulting schemes on modern parallel hardware. Benchmarking on a nonlinear hydrogen fuel cell model demonstrates order-of-magnitude speedups (by as much as three orders of magnitude) compared to sequential implementations.
\end{abstract}

\maketitle

\section{Introduction}

Constraint management is a central challenge in modern control systems. Safety, performance, and operational limits are often defined by hard constraints on states and inputs, and it is important to design control strategies that can guarantee these constraints are never violated.

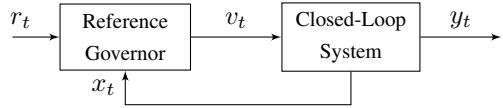
\begin{figure}

\centering
\begin{tikzpicture}[auto, node distance=1.5cm,>=latex']
\node [input, name=rinput] (rinput) {};
\node [block, right of=rinput,text width=1.5cm,align=center] (controller) {{\footnotesize Reference Governor}};
\node [block, right of=controller,node distance=3cm,text width=1.6cm,align=center] (system)
{{\footnotesize Closed-Loop System}};
\node [output, right of=system, node distance=2cm] (output) {};
\node [tmp, below of=controller,node distance=0.9cm] (tmp1){$s$};
\draw [->] (rinput) -- node{\hspace{-0.4cm}$r_t$} (controller);
\draw [->] (controller) -- node [name=v]{$v_t$}(system);
\draw [->] (system) -- node [name=y] {$y_t$}(output);
\draw [->] (system) |- (tmp1)-| node[pos=0.75] {$x_t$} (controller);
\end{tikzpicture}
\caption{Reference Governor block diagram. The variable $t$ is the discrete time index, $r_t$ is the desired setpoint at timestep $t$, $v_t$ is the modified setpoint, $y_t$ is the constrained output, and ${x}_t$ is the estimated state (or the true state, if available). The goal of RG is to maintain $y_t \in \mathbb{Y}$ for a given constraint set $\mathbb{Y}$.}
\label{fig:rg}
\end{figure}

Reference Governors (RGs) offer an elegant and computationally efficient framework for enforcing such constraints. Acting as supervisory add-ons to pre-stabilized feedback systems, RGs adjust the commanded reference signal so that the state and input constraints are satisfied while at the same time the properties of the underlying controller are retained when there is no risk of constraint violation. See Fig. \ref{fig:rg} for a block diagram of the RG setup.  The RG algorithm has several attractive properties: under mild assumptions, it is guaranteed to be recursively feasible, it can be made robust to modeling uncertainties and disturbances, and for linear closed-loop systems, it requires only a small computational footprint, making it suitable for embedded applications. 

Comprehensive surveys of RG  are available in \cite{garone2017reference,kolmanovsky2014reference}. The RG algorithm first appeared in \cite{194356, 83532, gilbert1995discrete} for linear systems, and was later extended to nonlinear systems in works such as \cite{kalabic2015reference, franze2014reconfigurable,vahidi2006constraint,angeli1999command,gilbert2002nonlinear,osorio2022novel,bemporad1998reference,sun2005load,ayubirad2025machine}. 
In particular,  \cite{bemporad1998reference,sun2005load,ayubirad2025machine} proposed a bisectional search combined with predictive online numerical simulations of the system dynamics to find an optimal constraint-admissible reference. This procedure can be computationally demanding due to the large number of simulations required.  The challenge becomes even more severe for systems with uncertainties, where robust and stochastic formulations of RG become computationally intractable. 


As it turns out, however, many of the required computations are naturally parallelizable, and parallel computing has already transformed related control algorithms such as real-time MPC \cite{abughalieh2019survey,bishop2024relu,yu2017efficient} and fast trajectory optimization \cite{heinrich2015real,rastgar2021gpu}. Multi-core Central Processing Units (CPUs) can provide modest speedups through a handful of compute cores, while Graphical Processing Units (GPUs) offer massive concurrency with thousands of lightweight threads \cite{tokhi2003parallel}. In this paper, we  investigate parallel implementations of the RG algorithm to unleash their full potential. 

We first formulate a Monte Carlo-based robust RG algorithm for nonlinear systems (Section~\ref{sec:RG}). The algorithm involves simulating the system dynamics under sampled uncertainties and disturbances from a known distribution (i.e., scenarios) and checking for constraint violations within a prediction horizon, similar in spirit to Scenario MPC  \cite{schildbach2014scenario}. To obtain probabilistic guarantees of constraint enforcement, the proposed robust RG may require thousands of disturbance scenarios, which renders the algorithm intractable on a single compute core -- a key reason why such formulations have not received much attention in the literature. To overcome this issue, we analyze the computational structure of the proposed robust RG algorithm, identify opportunities for parallel execution, and present implementations that exploit multi-core CPUs and CUDA-enabled GPUs (Section \ref{sec:parallel}). We then benchmark sequential and parallel implementations of the algorithm on a representative example from our past work, namely an automotive hydrogen fuel cell simulation \cite{ayubirad2025machine}, to quantify speedups and trade-offs (Section~\ref{sec:performance}). Our results highlight computational performance improvements of up to three orders of magnitude on a GPU as compared to a serial CPU implementation. 

In summary, the contributions of this paper are threefold:
\begin{itemize}
    \item A scenario-based robust RG algorithm for uncertain nonlinear systems;  
    \item Development of parallel implementation strategies for the proposed robust RG algorithm;
    \item A comparative analysis of CPU- and GPU-based implementations of the proposed parallel algorithm.
    \end{itemize}
        
    By bridging the gap between reference governor theory and high-performance computing, this work shows that robust RG formulations, which have been regarded as computationally intractable, are now within the realm of real-time implementation.



\section{Reference Governors for Constraint Management of Dynamical Systems}\label{sec:RG}

This section reviews the RG framework for nonlinear systems and then extends it to the robust case with disturbances. For detailed derivations of the non-robust case, see \cite{ayubirad2025machine}.  

\subsection{Reference governor background}

Consider the block diagram in Fig. \ref{fig:rg}, where the closed-loop system is described by:
\begin{equation}\label{eq:system}
\begin{gathered}
x_{t+1}=f(x_t,v_t)\\
y_t=h(x_t,v_t)
\end{gathered}
\end{equation}
where $x_t \in \mathbb{R}^n$ is the state, $v_t$ is the setpoint, and $y_t$ is the output to be constrained. The functions $f$ and $h$ are Lipschitz continuous.
For simplicity, we assume that the system is single-input, single-output (i.e., $u_t, y_t \in \mathbb{R}$) and that the state is accurately measured. Furthermore, we assume that the closed-loop system is stable (e.g., pre-stabilized by an inner-loop controller). 

Over the output, we impose the constraint: 
$$
y_t\in \mathbb{Y}
$$
where $\mathbb{Y}$ is non-empty and convex with 0 in its interior. The RG computes $v_t$ at each time step to enforce $y_t\in\mathbb{Y}$ for all future times. This is achieved by parameterizing the update as
\begin{equation}\label{eq:v}
v_t = v_{t-1} + \kappa (r_t - v_{t-1}), \qquad \kappa \in [0,1],
\end{equation}
so that $v_t$ is a convex combination of the previous input and the current desired setpoint $r_t$. Maximizing $\kappa$ drives $v_t$ as close as possible to $r_t$. 

Because constraints (i.e., $y_t\in\mathbb{Y}$) must hold for all future timesteps, directly enforcing these constraints would require an infinitely many inequalities, which is impossible to enforce on a computer. 
To overcome this, a tightened version of the constraint is enforced for the steady-state output: 
$$
y_{ss} \triangleq \lim_{j \rightarrow \infty} y_j \in (1-\epsilon) \mathbb{Y}
$$
where $\epsilon\in(0,1)$ is chosen to be (ideally) small. Note that, thanks to our stability assumption, $\lim_{j \rightarrow \infty} y_j$ exists and is a function of the constant input $v$. If the closed-loop system is linear, $y_{ss} = H_0 v$, where $H_0$ is the DC-gain of the closed-loop system. If the system is nonlinear, then we have that $y_{ss}(v)$ is the equilibrium manifold, which we assume is known and available for computations. 

With the tightened steady-state constraint, it can be shown that there exists a timestep $j^*$ such that
$$
 y_{ss} \in (1-\epsilon) \mathbb{Y} \;\;\mathrm{and}\;\; y_j\in\mathbb{Y}, j=0,\ldots,j^* \; \Rightarrow \; y_j\in\mathbb{Y}, \; \forall j
$$
leading to a finite number of non-redundant inequalities. Note that $j^*$ depends on the choice of $\epsilon$. Typically, the smaller the $\epsilon$, the larger the $j^*$ and therefore $\epsilon$ determines the complexity of the optimization problem. Larger $\epsilon$ leads to a less complex optimization problem that potentially solves faster, but also leads to a more conservative closed-loop response.

Summarizing the above, the RG optimization problem is given by:
\begin{equation}\label{eq:optim}
\begin{aligned}
\kappa^* \triangleq \max_{\kappa,v,\hat{x}_{j}}\quad & \kappa\\
{\rm s.t.}\;\quad & \kappa\in[0,1]\\
 & v=v_{t-1}+\kappa(r_{t}-v_{t-1})\\
 & \hat{x}_{j+1}=f(\hat{x}_{j},v),\;\;\;j=0,\ldots,j^{*}-1\\
 & h(\hat{x}_{j},v)\in\mathbb{Y},\;\;\;j=0,\ldots,j^{*}\\
 & y_{ss}(v)\in(1-\epsilon)\mathbb{Y}\\
 & \hat{x}_{0}=x_{t}
\end{aligned}
\end{equation}
where $\hat{x}_j$ are predicted states under constant $v$. In this optimization problem, $f$, $h$, $\mathbb{Y}$, $\epsilon$, and $j^*$ are known functions/parameters that do not change from one timestep to the next, while ${x}_t$, $v_{t-1}$, and $r_t$ are input parameters that do change. The decision variables are $\kappa$, $v$, and $\hat{x}_j$, with $v$ being the input that is eventually applied to the system. The optimization problem above is recursively feasible as long as it is feasible in the first timestep.

\subsection{The Reference Governor Algorithm}

Problem \eqref{eq:optim} is nonlinear and generally nonconvex. However, since $v$ depends affinely on the scalar $\kappa$, the optimization reduces to a one-dimensional search over $\kappa \in [0,1]$. This search can be carried out using a bisection algorithm, as proposed in \cite{bemporad1998reference,sun2005load} and summarized in Algorithm~1. The algorithm repeatedly simulates the dynamics for $j^*$ timesteps under different candidate values of $\kappa$ (and hence different constant inputs $v$), and identifies the largest feasible $\kappa$, denoted by $\kappa_{\mathrm{opt}}$, that satisfies the constraints. The parameter $N_\kappa$ controls the number of bisection iterations and guarantees that the computed solution is within  $0.5^{N_\kappa}$ of the true optimum, $k^*$. 


\begin{algorithm}
\caption{The nonlinear RG algorithm to compute \( \kappa \) (non-parallelized, non-robust; based on bisectional search)}
\label{alg:kappa_opt}
\begin{algorithmic}[1]
\STATE \textbf{Input:} $x_t$, $v_{t-1}$, $r_t$
\STATE \textbf{Output:} $\kappa_{\text{opt}}$
\STATE Initialize $\underline{\kappa} \gets 0$, $\overline{\kappa} \gets 1$, $\kappa \gets 1$, $\kappa_{\text{opt}} \gets 0$
\FOR{$i = 1$ to $N_{\kappa}$}
    \STATE $v \gets v_{t-1} + \kappa(r_t - v_{t-1})$
    \STATE Compute the predictions $\hat{x}_j$ and $y_j$, $j = 0, \ldots, j^*$ starting from $x_t$ and constant input $v$ from line 5.
   
    \IF{$y_j \in \mathbb{Y}$  for all timesteps and ${y}_{ss}(v) \in (1-\epsilon) \mathbb{Y}$}
        \STATE $\kappa_{\text{opt}} \gets \kappa$
        \IF{$\kappa = 1$}
            \STATE \textbf{break}
        \ELSE
            \STATE $\underline{\kappa} \gets \kappa$, $\kappa \gets \left(\underline{\kappa} + \overline{\kappa}\right) / 2$
        \ENDIF
    \ELSE
        \STATE $\overline{\kappa} \gets \kappa$, $\kappa \gets \left(\underline{\kappa} + \overline{\kappa}\right) / 2$
    \ENDIF
\ENDFOR
\end{algorithmic}\label{alg:PRG}
\end{algorithm}

\subsection{Robust RG Extension}

In the presence of uncertainties and unknown external disturbances, model \eqref{eq:system} becomes:
\begin{equation}\label{eq:system2}
\begin{gathered}
x_{t+1}=f(x_t,v_t) + d_t\\
y_t=h(x_t,v_t)
\end{gathered}
\end{equation}
where $d_t \in \mathbb{R}^n$. Two common modeling approaches for $d_t$ are: ($i$) set-bounded (e.g., \cite{480258}), i.e.,  $d_t\in\mathbb{D}$. In this case, the RG must enforce $y_t\in\mathbb{Y}$ for all possible realizations of $d_t$, including the worst case. And ($ii$) stochastic (e.g., \cite{KALABIC2019108500, 8511439}), where $d_t$ is modeled as a random variable with a known distribution. Here, the RG must enforce $y_t\in\mathbb{Y}$ with high probability (chance constraints).

Robust RG formulations are well developed for linear systems, but tractable extensions to nonlinear are not available. To address this gap, we propose a scenario-based approach below.

Let $N_{\text{sim}}$ denote the number of disturbance scenarios considered. At each timestep, we draw $N_{\text{sim}}$ disturbance sequences from the probability distribution of $d_t$, yielding realizations $d_j^k$, $j=0,\ldots,j^*$, $k=1,\ldots,N_{\text{sim}}$, where the subscript denotes the timestep and the superscript denotes the scenario index. For each scenario $k$, the system dynamics are simulated forward under $d_j^k$ with various values of $\kappa$ (similar to Algorithm 1) to determine the largest $\kappa$ that satisfies the constraints over the prediction horizon. The final update is chosen as the minimum of these $\kappa$ values across all sampled scenarios to ensure robustness. The resulting algorithm is summarized in Algorithm~\ref{alg:PRG22}.

Note that with persistent time-varying disturbances, the output under a constant input 
$v$ does not converge to an equilibrium point but rather to a bounded tube around the disturbance-free steady-state $y_{ss}(v)$. The parameter 
$\epsilon$ in Algorithm 1 (step 6 of Algorithm 2) can be chosen large enough to contain this tube, which yields a surrogate that may be potentially conservative. 

From the perspective of set-bounded disturbance modeling introduced above, the scenario-based strategy described here replaces the intractable universal quantifier (``for all realizations of $d_t$'') with a finite sampled set. From the perspective of stochastic modeling, it corresponds to the classical scenario method, for which established results provide probabilistic guarantees of feasibility as a function of the number of samples (see, e.g., \cite{schildbach2014scenario}).

\begin{remark}
   The theoretical properties of the proposed RG, including probabilistic recursive feasibility, are difficult to establish rigorously in the stochastic case and will be the subject of future work. Nevertheless, numerical simulations in Section IV show that with sufficiently large $N_{\text{sim}}$, $j^*$, and $\epsilon$, the algorithm remains recursively feasible.
\end{remark}



\begin{algorithm}
\caption{The nonlinear RG algorithm to compute \( \kappa \) (non-parallelized, robust; based on bisectional search)}
\label{alg:kappa_optnonparalleldist}
\begin{algorithmic}[1]
\STATE \textbf{Input:} $x_t$, $v_{t-1}$, $r_t$
\STATE \textbf{Output:} $\kappa_{\text{opt}}$
\STATE $\kappa_{\mathrm{opt}} \gets 1$
\FOR{$k = 1$ to $N_{\text{sim}}$}
\STATE Sample a disturbance sequence $d_j^k$, $j=0,\ldots,j^*$. 
\STATE Run Algorithm~1, but in each forward simulation step use $\hat{x}_{j+1}=f(\hat{x}_j,v)+d_j^k$. Let the returned value be $\kappa_k$.
\STATE $\kappa_{\mathrm{opt}} \gets \min(\kappa_{\mathrm{opt}},\,\kappa_k)$
\ENDFOR
\end{algorithmic}\label{alg:PRG22}
\end{algorithm}

Note that Algorithm 1 can be viewed as a special case of Algorithm 2 with $d_t = 0$ and $N_{\text{sim}} = 1$. For this reason, in the next section we focus on parallelizing Algorithm 2.

\section{Parallelized Reference Governor}\label{sec:parallel}

\subsection{Parallelized algorithm}

In this section, we show that the robust reference governor can be parallelized. As it turns out, the bisectional search for $\kappa$ in Algorithms 1 and 2 is no longer suitable here, since each bisection iteration depends on the outcome of the previous one and thus prevents parallel execution. Instead, we adopt a grid search that allows for parallelization. Specifically, we discretize the interval 
$[0,1]$ into 
$M$ candidate values of 
$\kappa$. At each candidate value, the system dynamics are simulated for 
$N_\text{sim}$ different disturbance scenarios. There are thus two parallelization opportunities: (i) across the grid search, and (ii) across the disturbance scenarios. As we explain below, we exploit both opportunities.
 
As before, let the  disturbance realizations be denoted by $d_j^k$, $j=0,\ldots,j^*$, $k=1,\ldots,N_{\text{sim}}$. Suppose we have ${M \cdot N_\textrm{sim}}$ compute cores available, where $M$ is, as before, the grid size for $\kappa$. Consider the cores to be arranged conceptually into an $M \times N_\textrm{sim}$ array. Each core independently executes the  the non-parallelized steps of the RG algorithm for one disturbance scenario and one $\kappa$ value: it returns the value 1 if the simulation satisfies the constraints for all $j^*+1$ timesteps, and 0 if the constraints are violated at any timestep. The returned results are collected into a 2-D boolean array $P \in \{0,1\}^{M\times N_{\text{sim}}}$. Once all of the simulations have terminated, one compute core scans $P$ to identify the maximum row index $i^*$ such that $P_{i^*k} = 1$ for all $k$. This index is mapped to the corresponding $\kappa$ value, yielding $\kappa_\textrm{opt}$:
\begin{equation}
\begin{aligned}
\kappa_{{\rm opt}}=\max_{i\in\{1,2,...M\}}\quad & \frac{i-1}{M-1}\\
{\rm s.t.}\quad & P_{ik}=1,\forall k
\end{aligned}
\end{equation}
This parallelized algorithm is summarized in Algorithm~3. 

\begin{algorithm}
\caption{The nonlinear RG algorithm to compute \( \kappa \) (parallelized, robust, based on grid search)}
\label{alg:kappa_optparalleldist}
\begin{algorithmic}[1]
\STATE \textbf{Input:} $x_t$, $v_{t-1}$, $r_t$
\STATE \textbf{Output:} $\kappa_{\text{opt}}$
  \STATE Grid the interval $[0,1]$ into $M$ points: $\kappa_i = (i-1) \frac{1}{M-1}$, $i=1, \ldots, M$. 
  \STATE Draw $N_{\text{sim}}$ realizations of the disturbance: $d_j^k$, $j=0,\ldots,j^*$, $k=1,\ldots,N_{\text{sim}}$. 
    \STATE Each compute core with index $(i,k)$ performs the following calculations using $\kappa_i$ and $d^k_j$:
    \begin{algsubsteps}
        \STATE Parallel step 1: $v \gets v_{t-1} + \kappa_i (r_t - v_{t-1})$
        \STATE Parallel step 2:  
    Compute the predictions $\hat{x}$ and $y_j$, $j = 0, \ldots, j^*$ starting from $x_t$ and constant input $v$, with disturbance realization $d_{j}^k$.
    \STATE Parallel step 3: If $y_j \in \mathbb{Y}, j=0,\ldots, j^*$ and ${y}_{ss}(v) \in (1-\epsilon) \mathbb{Y}$, return $P_{ik} = 1$, otherwise  $P_{ik} = 0$.
    \end{algsubsteps}
    \STATE Let $i^* = \max \{i: P_{ik}=1 \text{ for all } k\}$. Return $k_{\text{opt}} = \frac{(i^*-1)}{M-1}$
\end{algorithmic}\label{alg:PRG2}
\end{algorithm}

\subsection{CPU parallelization}

To implement the parallel robust reference governor for the CPU, Algorithm 3 can almost be directly translated into code. The primary difference is that while Algorithm~3 assumes as many cores as tasks, in practice CPUs have far fewer cores than the assumed $M \cdot N_{\text{sim}}$. Modern operating systems can run programs on more threads than the hardware has physical cores by context switching, but each additional thread introduces some overhead. Because of this overhead it makes sense for the program to run only as many threads as there are cores. Instead of each simulation running on a dedicated thread, each thread runs all of the simulations for some subset of the $M$ $\kappa$ values being tested at each time step. The simulation constraints are checked at each discrete time step, and simulations with states that fall outside the constraints are terminated. Each thread allocates a block of memory for local simulation, which remains allocated while it is running. All threads write the results of their simulations to a shared memory array. Synchronization is not required for these writes, as two threads will never attempt to write results for the same simulation. Once all the simulation threads are complete, the main thread searches the output array for the highest valid $\kappa$ value.

\subsection{GPU parallelization}

Algorithm 3 maps even more directly into GPU code. Because the GPU has significantly more compute cores than the CPU, and the GPU scheduler is designed to efficiently run many more threads than there are cores, each simulation gets mapped to a single thread. There is still a limit to the number of threads that can execute simultaneously, which depends on the specific GPU hardware being used. The threads are grouped into $16 \times 16$ blocks with the number of blocks depending on the granularity of the $\kappa$ grid and the number of disturbance scenarios. The output array, disturbances, and system parameters are stored in global memory on the GPU. Faster block-level shared memory is allocated for each thread to store intermediate states of the local simulation. Like on the CPU, individual threads terminate as soon as the constraints are breached. Unlike on the CPU, the GPU compute cores are only capable of fast operations on single precision floating point numbers. Using all single precision floating point numbers does result in less accuracy, but it allows for a significant performance increase (the loss of accuracy had no visible effect in our numerical simulations). Once all threads are completed, the output array is transferred back to host (CPU-accessible) memory. This memory transfer is a significant bottleneck for the program, which will be seen in the performance figures in the next section. The main CPU thread then performs the same search for valid $\kappa$ values as in the CPU implementation. Note that, in the GPU implementation of the RG, the CPU is still responsible for host-side computations and for initiating GPU kernels.

\section{Performance Analysis}\label{sec:performance}

In this section, we compare CPU and GPU parallelization of the robust RG algorithm (Algorithm 3) on an automotive hydrogen fuel cell model from our prior work \cite{ayubirad2025machine,ayubirad2025neural}.

\subsection{Fuel cell model}

Fuel cell (FC) systems are devices that generate electricity via electrochemical reactions, typically between hydrogen and oxygen. The FC model used in this section is a reduced third-order nonlinear, spatially averaged, model of a FC stack with its auxiliaries \cite{talj2009experimental}. In this model, the fuel cell stack consists of $381$ cells and can supply up to $300$ A of current. This low-order model is derived from a full-order system developed in \cite{pukrushpan2004control}, and is suitable for studying the air-path side of the FC system. The states of the model are defined as $x={\left[p_{ca},{\omega }_{cp},p_{sm}\right]}^\top$, consisting of the cathode air pressure, the compressor motor speed, and the supply manifold pressure, respectively. The airpath system contains a low-level controller, which controls the compressor to maximize net power output. The details of the model and the feedforward and feedback controller design can be found in \cite{ayubirad2025machine,ayubirad2025neural}. 

The constrained output of the air-path subsystem is the so-called  Oxygen Excess Ratio (OER). The goal is to maintain the OER above 1.9 to prevent oxygen starvation, i.e.,
$$
y_t \ge 1.9
$$
During a typical step-up of current demand, the OER constraint may be violated. A reference governor is thus employed to enforce the OER constraint.

\begin{figure}
    \centering
    \includegraphics[width=\linewidth]{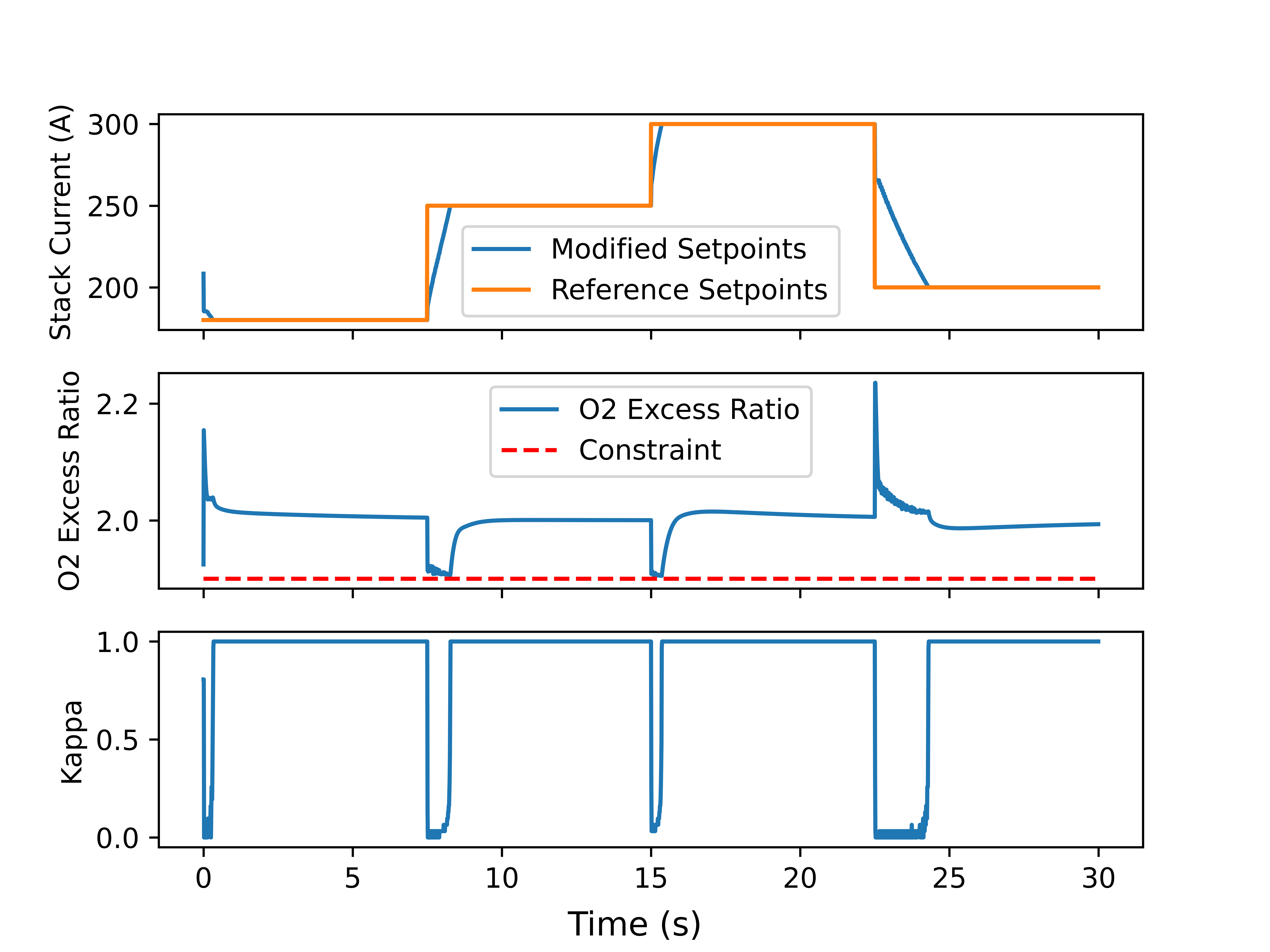}
    \caption{Fuel Cell Simulation}
    \label{fig:sim}
\end{figure}

To this end, the model of the closed-loop system is discretized using a 4th order Runge Kutta algorithm with a fixed step size of 0.01s. The simulation results of the discretized system with and without a robust reference governor (Algorithm 3) are shown in Fig. \ref{fig:sim}. The following parameters were used to generate this simulation: $j^*=1024$, $N_{\text{sim}}=8192$, and the disturbance distributions are all uniform, with ranges $\pm 50$, $\pm 10$, and $\pm 50$ for the first, second, and third states, respectively. As can be seen in the figure, the reference governor enforces the constraint for all time. 

\subsection{Computational performance of the parallelized algorithm}

To test the performance of the parallel reference governor, Algorithm 3 was implemented in C++. The code used to generate the simulation in Fig. \ref{fig:sim} and the benchmarking results described below can be found on our GitHub repository: \url{https://github.com/wshayne/RefGovernor-Public}.

Performance analysis was conducted on two computers: a desktop windows PC with 64GB of 4800 MT/s DDR5 RAM, an RTX 3090 GPU with 24GB of GDDR6X VRAM and 10,496 CUDA cores, and Ryzen 9 7950x3d CPU with 16 cores (base clock 4.2 GHz, boost clock 5.7GHz); and a laptop with 16GB of RAM, an RTX 4070 (mobile) GPU with 8GB of GDDR6 VRAM and 4608 CUDA cores, and an i7 13700H CPU with 6 performance cores (base clock 2.4 GHz, boost clock 5.0GHz) and 8 efficiency cores (base clock 1.8GHz, boost clock 3.7 GHz). The number of disturbance scenarios (i.e., $N_{\text{sim}}$) tested ranged from 1 to 8192, and the grid width of $\kappa$ (i.e., $M$) was set at 32. Disturbance counts from 1 to 32 were all tested, and disturbance counts from 32 to 8192 were tested in steps of 32. To reduce the impact of background processes on the computer on the timing of the algorithm, tests of disturbance counts from 1-32 are averaged over 200 runs. GPU results for 32-8192 disturbances are also averaged over 200 runs. Parallel CPU results for 32-8192 disturbances are averaged over 22 runs since they took much longer to complete. Single thread CPU results for 32-8192 disturbances are averaged over 3 runs. Two benchmarks are recorded for the GPU, one time including the generation of random disturbances and transfers of parameters to and results from the GPU memory and a second only including the runtime of the algorithm.

\begin{figure}
    \centering
    \includegraphics[width=\linewidth]{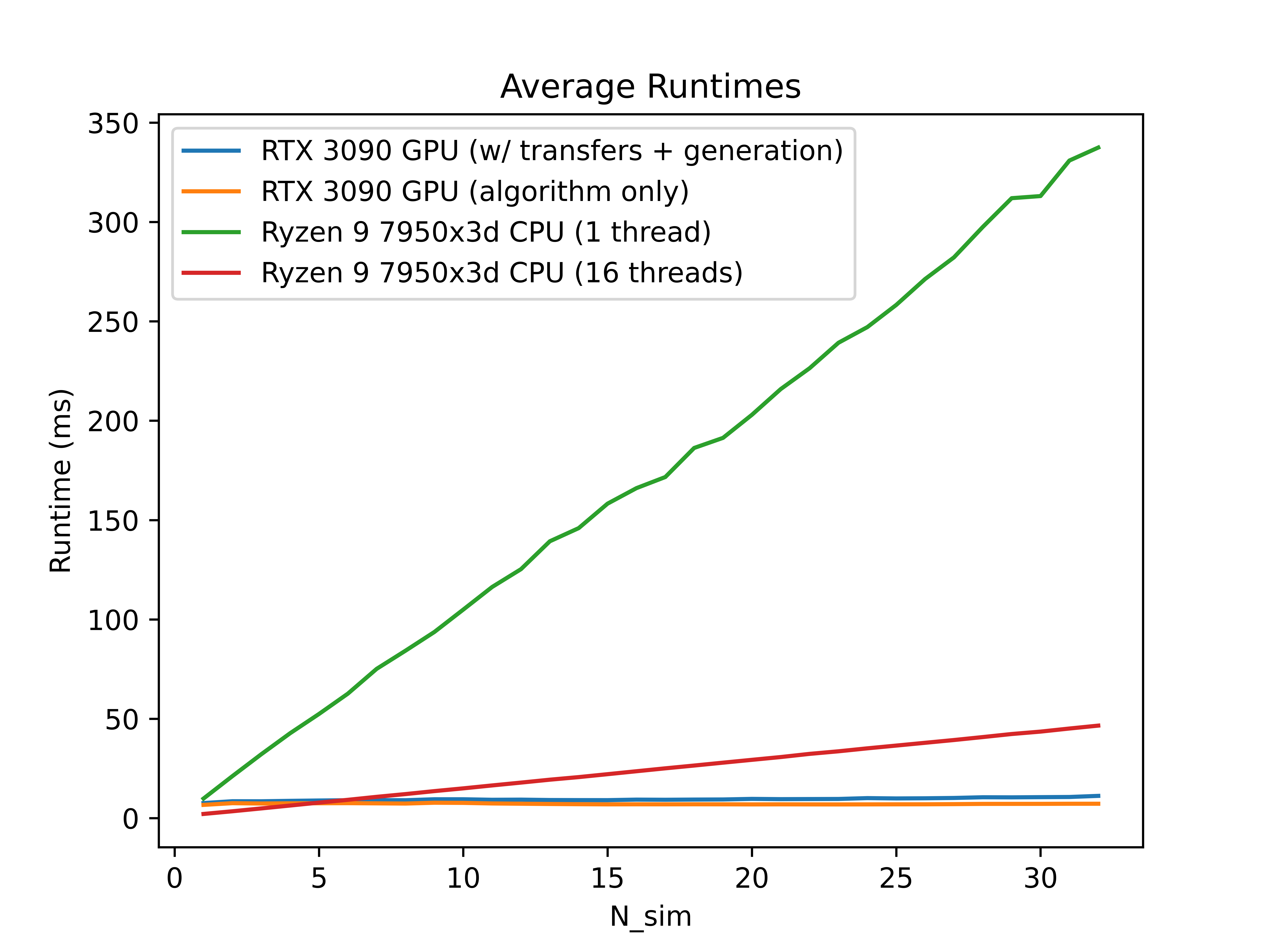}
    \caption{Desktop PC Benchmark (1-32 Disturbances). RTX traces denote GPU implementation; Ryzen traces denote CPU implementations. $N_{\text{sim}}$ on the x-axis denotes the number of disturbance scenarios considered.}
    \label{fig:desktop_lowcount}
\end{figure}

Figure \ref{fig:desktop_lowcount} shows how quickly the runtime for the CPU serial and parallelized versions of the algorithm increase, while the performance of the GPU algorithm remains effectively constant for up to 32 disturbances. The GPU parallelized version of the algorithm for this problem with a $\kappa$ search width of 32 is faster than the CPU parallelized version when working with more than 5 disturbances.

\begin{figure}
    \centering
    \includegraphics[width=\linewidth]{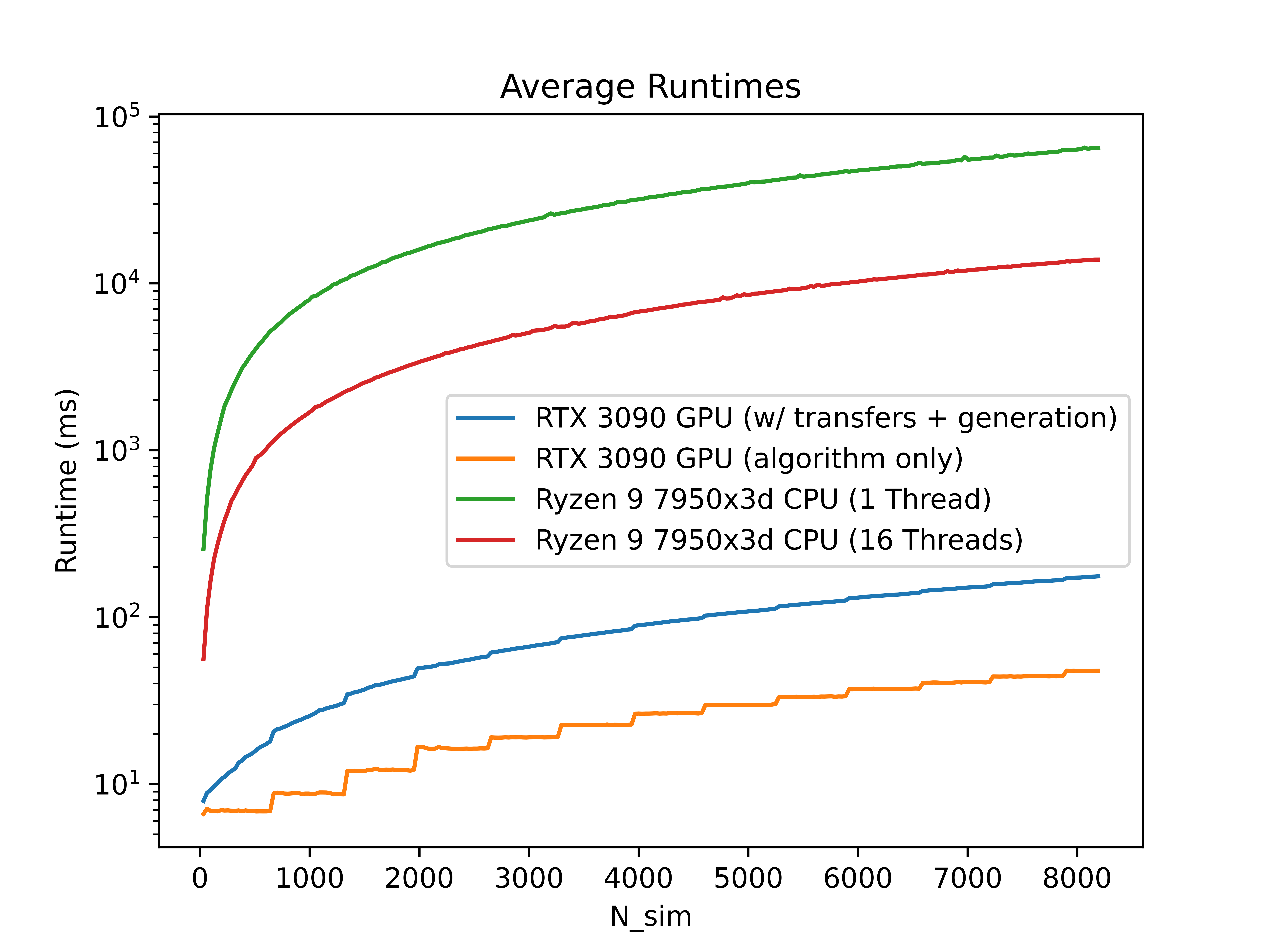}
    \caption{Desktop PC Benchmark (32-8192 Disturbances). RTX traces denote GPU implementation; Ryzen traces denote CPU implementations.}
    \label{fig:desktop_benchmark}
\end{figure}

Figure \ref{fig:desktop_benchmark} shows the overall performance curve for both the CPU and GPU implementations of the algorithm at higher disturbance counts. Note that the y-axis is logarithmic. The shape of the GPU curve without memory transfers is indicative of how resources are scheduled and used. Once the GPU runs out of additional resources to perform the simulations in parallel, the next batches of simulations get run in series. Each ``step'' up on the runtime curve represents an additional required batch. We can see that, in this case, a new batch is added every 600-700 disturbances. These exact numbers will change depending on the hardware used and the system being simulated, but the pattern will remain the same.

\begin{figure}
    \centering
    \includegraphics[width=1\linewidth]{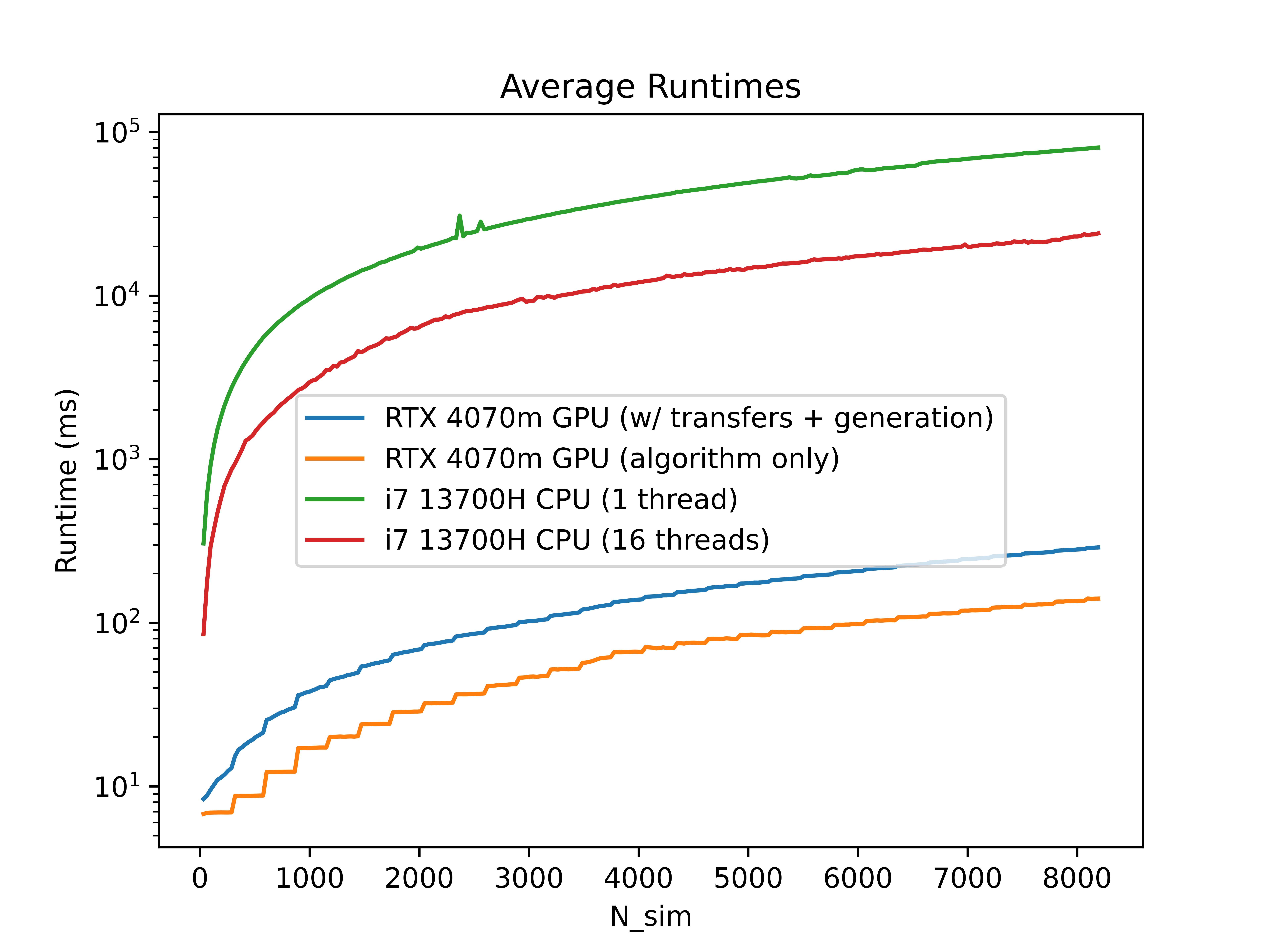}
    \caption{Laptop PC Benchmark (32-8192 Disturbances). RTX traces denote GPU implementation; i7 traces denote CPU implementations.}
    \label{fig:laptop_benchmark}
\end{figure}

The results on the laptop computer are shown in Fig. \ref{fig:laptop_benchmark}. For brevity, we only show the results for 32-8192 disturbances. As can be seen, the algorithm runs slower on the laptop overall, and the GPU hardware is less capable so a new batch is added roughly every 300 disturbances, but the overall shapes of the curves remain the same. Additional or more powerful GPUs could be utilized to attain greater performance, but even on the older consumer hardware we tested performance is good enough to provide control updates every 50ms at 512 disturbances, or every 100ms at 1024 disturbances. This indicates that the proposed robust RG algorithm is real-time feasible if parallel computing hardware is available. 

As a final remark, we note that when $N_{\text{sim}}=1$, parallelization does not yield a significant speedup. Thus, classical RG formulations, which correspond to this case, do not benefit from parallelization, while the robust formulation proposed here does.


\section{Conclusions and Future Work}

This paper presented a robust reference governor (RG) formulation for nonlinear systems and investigated its parallel implementation on multi-core CPUs and CUDA-enabled GPUs. 
By exploiting the inherent parallel structure of RG, notably the repeated forward simulations and constraint checks, we demonstrated substantial reductions in computation time compared to sequential implementations. 
The results indicate that CPU parallelization scales well for moderate numbers of disturbance scenarios, while GPU acceleration offers dramatic speedups in large-scale settings, reaching up to three orders of magnitude faster than serial CPU implementations. The results presented here hold the potential to make robust RG practical for nonlinear systems.

Several challenges remain open. First, the theoretical properties of the scenario-based robust RG, including rigorous guarantees of probabilistic recursive feasibility, require further study. Second, numerical aspects of GPU implementations, including the impact of reduced-precision arithmetic, warrant further analysis. Finally, extending the framework to more complex applications, e.g., high-dimensional nonlinear plants and real-time embedded implementations, is an interesting topic for future research.



\bibliographystyle{IEEEtran}
\bibliography{refs} 

\begin{thebibliography}{10}
\providecommand{\url}[1]{#1}
\csname url@rmstyle\endcsname
\providecommand{\newblock}{\relax}
\providecommand{\bibinfo}[2]{#2}
\providecommand\BIBentrySTDinterwordspacing{\spaceskip=0pt\relax}
\providecommand\BIBentryALTinterwordstretchfactor{4}
\providecommand\BIBentryALTinterwordspacing{\spaceskip=\fontdimen2\font plus
\BIBentryALTinterwordstretchfactor\fontdimen3\font minus \fontdimen4\font\relax}
\providecommand\BIBforeignlanguage[2]{{%
\expandafter\ifx\csname l@#1\endcsname\relax
\typeout{** WARNING: IEEEtran.bst: No hyphenation pattern has been}%
\typeout{** loaded for the language `#1'. Using the pattern for}%
\typeout{** the default language instead.}%
\else
\language=\csname l@#1\endcsname
\fi
#2}}

\bibitem{garone2017reference}
E.~Garone, S.~Di~Cairano, and I.~Kolmanovsky, ``Reference and command governors for systems with constraints: A survey on theory and applications,'' \emph{Automatica}, vol.~75, pp. 306--328, 2017.

\bibitem{kolmanovsky2014reference}
I.~Kolmanovsky, E.~Garone, and S.~Di~Cairano, ``Reference and command governors: A tutorial on their theory and automotive applications,'' in \emph{2014 American Control Conference}.\hskip 1em plus 0.5em minus 0.4em\relax IEEE, 2014, pp. 226--241.

\bibitem{194356}
P.~Kapasouris, M.~Athans, and G.~Stein, ``Design of feedback control systems for stable plants with saturating actuators,'' in \emph{Proceedings of the 27th IEEE Conference on Decision and Control}, 1988, pp. 469--479 vol.1.

\bibitem{83532}
E.~Gilbert and K.~Tan, ``Linear systems with state and control constraints: the theory and application of maximal output admissible sets,'' \emph{IEEE Transactions on Automatic Control}, vol.~36, no.~9, pp. 1008--1020, 1991.

\bibitem{gilbert1995discrete}
E.~G. Gilbert, I.~Kolmanovsky, and K.~T. Tan, ``Discrete-time reference governors and the nonlinear control of systems with state and control constraints,'' \emph{International Journal of robust and nonlinear control}, vol.~5, no.~5, pp. 487--504, 1995.

\bibitem{kalabic2015reference}
U.~Kalabic, ``Reference governors: Theoretical extensions and practical applications,'' Ph.D. dissertation, University of Michigan, 2015.

\bibitem{franze2014reconfigurable}
G.~Franze, M.~Mattei, L.~Ollio, V.~Scordamaglia, and F.~Tedesco, ``A reconfigurable aircraft control scheme based on an hybrid command governor supervisory approach,'' in \emph{2014 American Control Conference}.\hskip 1em plus 0.5em minus 0.4em\relax IEEE, 2014, pp. 1273--1278.

\bibitem{vahidi2006constraint}
A.~Vahidi, I.~Kolmanovsky, and A.~Stefanopoulou, ``Constraint handling in a fuel cell system: A fast reference governor approach,'' \emph{IEEE Transactions on Control Systems Technology}, vol.~15, no.~1, pp. 86--98, 2006.

\bibitem{angeli1999command}
D.~Angeli and E.~Mosca, ``Command governors for constrained nonlinear systems,'' \emph{IEEE Transactions on Automatic Control}, vol.~44, no.~4, pp. 816--820, 1999.

\bibitem{gilbert2002nonlinear}
E.~Gilbert and I.~Kolmanovsky, ``Nonlinear tracking control in the presence of state and control constraints: a generalized reference governor,'' \emph{Automatica}, vol.~38, no.~12, pp. 2063--2073, 2002.

\bibitem{osorio2022novel}
J.~Osorio, M.~Santillo, J.~B. Seeds, M.~Jankovic, and H.~R. Ossareh, ``A novel reference governor approach for constraint management of nonlinear systems,'' \emph{Automatica}, vol. 146, p. 110554, 2022.

\bibitem{bemporad1998reference}
A.~Bemporad, ``Reference governor for constrained nonlinear systems,'' \emph{IEEE Transactions on Automatic Control}, vol.~43, no.~3, pp. 415--419, 1998.

\bibitem{sun2005load}
J.~Sun and I.~V. Kolmanovsky, ``Load governor for fuel cell oxygen starvation protection: A robust nonlinear reference governor approach,'' \emph{IEEE Transactions on Control Systems Technology}, vol.~13, no.~6, pp. 911--920, 2005.

\bibitem{ayubirad2025machine}
M.~Ayubirad and H.~R. Ossareh, ``A machine learning-based reference governor for nonlinear systems with application to automotive fuel cells,'' \emph{IEEE Transactions on Control Systems Technology}, 2025.

\bibitem{abughalieh2019survey}
K.~M. Abughalieh and S.~G. Alawneh, ``A survey of parallel implementations for model predictive control,'' \emph{IEEE Access}, vol.~7, pp. 34\,348--34\,360, 2019.

\bibitem{bishop2024relu}
A.~L. Bishop, J.~Z. Zhang, S.~Gurumurthy, K.~Tracy, and Z.~Manchester, ``Relu-qp: A gpu-accelerated quadratic programming solver for model-predictive control,'' in \emph{2024 IEEE International Conference on Robotics and Automation (ICRA)}.\hskip 1em plus 0.5em minus 0.4em\relax IEEE, 2024, pp. 13\,285--13\,292.

\bibitem{yu2017efficient}
L.~Yu, A.~Goldsmith, and S.~Di~Cairano, ``Efficient convex optimization on gpus for embedded model predictive control,'' in \emph{Proceedings of the general purpose GPUs}, 2017, pp. 12--21.

\bibitem{heinrich2015real}
S.~Heinrich, A.~Zoufahl, and R.~Rojas, ``Real-time trajectory optimization under motion uncertainty using a gpu,'' in \emph{2015 IEEE/RSJ International Conference on Intelligent Robots and Systems (IROS)}.\hskip 1em plus 0.5em minus 0.4em\relax IEEE, 2015, pp. 3572--3577.

\bibitem{rastgar2021gpu}
F.~Rastgar, H.~Masnavi, J.~Shrestha, K.~Kruusam{\"a}e, A.~Aabloo, and A.~K. Singh, ``Gpu accelerated convex approximations for fast multi-agent trajectory optimization,'' \emph{IEEE Robotics and Automation Letters}, vol.~6, no.~2, pp. 3303--3310, 2021.

\bibitem{tokhi2003parallel}
M.~O. Tokhi, M.~A. Hossain, and M.~H. Shaheed, \emph{Parallel computing for real-time signal processing and control}.\hskip 1em plus 0.5em minus 0.4em\relax Springer Science \& Business Media, 2003.

\bibitem{schildbach2014scenario}
G.~Schildbach, L.~Fagiano, C.~Frei, and M.~Morari, ``The scenario approach for stochastic model predictive control with bounds on closed-loop constraint violations,'' \emph{Automatica}, vol.~50, no.~12, pp. 3009--3018, 2014.

\bibitem{480258}
E.~Gilbert and I.~Kolmanovsky, ``Discrete-time reference governors for systems with state and control constraints and disturbance inputs,'' in \emph{Proceedings of 1995 34th IEEE Conference on Decision and Control}, vol.~2, 1995, pp. 1189--1194 vol.2.

\bibitem{KALABIC2019108500}
\BIBentryALTinterwordspacing
U.~V. Kalabić, N.~I. Li, C.~Vermillion, and I.~V. Kolmanovsky, ``Reference governors for chance-constrained systems,'' \emph{Automatica}, vol. 109, p. 108500, 2019. [Online]. Available: \url{https://www.sciencedirect.com/science/article/pii/S0005109819303619}
\BIBentrySTDinterwordspacing

\bibitem{8511439}
J.~Osorio and H.~R. Ossareh, ``A stochastic approach to maximal output admissible sets and reference governors,'' in \emph{2018 IEEE Conference on Control Technology and Applications (CCTA)}, 2018, pp. 704--709.

\bibitem{ayubirad2025neural}
M.~Ayubirad and H.~R. Ossareh, ``A neural network-based multi-timestep command governor for nonlinear systems with constraints,'' in \emph{2025 IEEE Conference on Control Technology and Applications (CCTA)}.\hskip 1em plus 0.5em minus 0.4em\relax IEEE, 2025, pp. 1--8.

\bibitem{talj2009experimental}
R.~J. Talj, D.~Hissel, R.~Ortega, M.~Becherif, and M.~Hilairet, ``Experimental validation of a pem fuel-cell reduced-order model and a moto-compressor higher order sliding-mode control,'' \emph{IEEE Transactions on Industrial Electronics}, vol.~57, no.~6, pp. 1906--1913, 2009.

\bibitem{pukrushpan2004control}
J.~T. Pukrushpan, A.~G. Stefanopoulou, and H.~Peng, \emph{Control of fuel cell power systems: principles, modeling, analysis and feedback design}.\hskip 1em plus 0.5em minus 0.4em\relax Springer Science \& Business Media, 2004.

\end{thebibliography}

\end{document}